\documentclass[conference]{IEEEtran}
\IEEEoverridecommandlockouts
\usepackage{cite}
\usepackage{amsmath,amssymb,amsfonts}
\usepackage{algorithmic}
\usepackage{graphicx}
\usepackage{verbatim}
\usepackage{textcomp}
\usepackage{xcolor}
\def\BibTeX{{\rm B\kern-.05em{\sc i\kern-.025em b}\kern-.08em
    T\kern-.1667em\lower.7ex\hbox{E}\kern-.125emX}}
\usepackage{caption}
\usepackage{subcaption}
\usepackage{multirow}
\usepackage{url}
\usepackage{hyperref}
\begin{document}

\title{Exploring Error Bits for Memory Failure Prediction: An In-Depth Correlative Study\vspace{-0.3cm}}

\author{

\IEEEauthorblockN{Qiao Yu\IEEEauthorrefmark{1}\IEEEauthorrefmark{2},
Wengui Zhang\IEEEauthorrefmark{3},
Jorge Cardoso\IEEEauthorrefmark{1}\IEEEauthorrefmark{5}, and
Odej Kao\IEEEauthorrefmark{2}}

\IEEEauthorblockA{\IEEEauthorrefmark{1}Huawei Munich Research Center, Germany, \{qiao.yu, jorge.cardoso\}@huawei.com\\
}

\IEEEauthorblockA{\IEEEauthorrefmark{2}Technical University of Berlin, Germany, odej.kao@tu-berlin.de\\
}

\IEEEauthorblockA{\IEEEauthorrefmark{3}Huawei Technologies Co., Ltd, China, zhangwengui1@huawei.com \\
}

\IEEEauthorblockA{\IEEEauthorrefmark{5}CISUC, University of Coimbra, Portugal}

\vspace{-1.1cm}}

\maketitle
\thispagestyle{plain}
\pagestyle{plain}

\begin{abstract}
In large-scale datacenters, memory failure is a common cause of server crashes, with Uncorrectable Errors (UEs) being a major indicator of Dual Inline Memory Module (DIMM) defects. Existing approaches primarily focus on predicting UEs using Correctable Errors (CEs), without fully considering the information provided by error bits. However, error bit patterns have a strong correlation with the occurrence of UEs. In this paper, we present a comprehensive study on the correlation between CEs and UEs, specifically emphasizing the importance of spatio-temporal error bit information. Our analysis reveals a strong correlation between spatio-temporal error bits and UE occurrence. Through evaluations using real-world datasets, we demonstrate that our approach significantly improves prediction performance by 15\% in F1-score compared to the state-of-the-art algorithms. Overall, our approach effectively reduces the number of virtual machine interruptions caused by UEs by approximately 59\%.

\end{abstract}


\section{Introduction}
\begin{figure}[t]
\centering
\includegraphics[keepaspectratio]{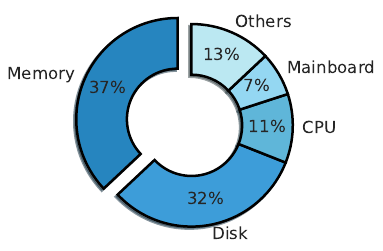}

\caption{Distribution of Hardware Failures in Data Centers \cite{JD_HW}.}
\vspace{-0.5cm}
\label{fig:distribution_HW}

\end{figure}

With the increasing demand for cloud computing and big data storage services, hardware failures \cite{hw_failure,optical_failure} can significantly impact the Reliability, Availability, and Serviceability (RAS)\footnote{Reliability, Availability, and Serviceability are three key attributes assessing the dependability of computer systems.} of servers. Among hardware failures, DRAM (Dynamic Random Access Memory) failure is a major occurrence, accounting for 37\% of total hardware failures in Figure~\ref{fig:distribution_HW}. DRAM failure is often accompanied by DRAM errors, i.e, Correctable Error (CE) and Uncorrectable Error (UE). To mitigate DRAM failures, Error Correction Code (ECC) mechanisms such as SEC-DED \cite{SECDED}, Chipkill \cite{chipkill1997} and SDDC \cite{intel_sddc} are used to detect and correct data corruption errors. For example, Chipkill ECC can correct any erroneous data bits originating from a single DRAM chip. However, when erroneous data bits span across two or more chips, the error correction capability of Chipkill ECC becomes overwhelmed, often resulting in a system crash due to a UE. Moreover, the ECC on contemporary Intel platforms like Skylake and Cascade Lake servers is less robust compared to Chipkill ECC, making it vulnerable to certain error-bit patterns from a single chip \cite{Li_intel_bit_2022}. Thus, soly depending on ECC for DRAM reliability proves inadequate, with DRAM failures remaining a significant cause of system failures.

To improve memory reliability, several studies \cite{Schroeder_DRAMErrors_2009,Meza_RevisitingError_2015,Sridharan_studyDRAM_2012,Sridharan_memory_error_modern_system,vilas_systematic_study,patel2022case_study} have investigated the correlations between memory errors and failures, which forms the foundation of our work. Machine Learning (ML)-based techniques have been leveraged for DRAM failure prediction \cite{giurgiu_predicting_2017,du_memory_2018,du2020-intel,boixaderas_cost-aware_2020,Yu_drampakdd_2021,Du_nofreepredictor_2021,srds_in-Depth_MEM,fisrtCE_matter_TUB}, using CEs information from a large-scale datacenter to predict UEs. These studies have effectively utilized the spatial distribution of CEs to enhance DRAM failure prediction. Moreover, system-level workload indicators such as memory utilization, read and write have been applied for DRAM failure prediction in \cite{sun2019-alibaba1,queen_workload-ware_2019,alibaba_workload-aware_2021}. The experiments in \cite{alibaba_workload-aware_2021} have demonstrated that the workload metric is relatively less significant compared to other CE related features. In \cite{alibaba_2022_node_prediction}, CE storm (numerous CEs occurring in a short period) and UEs are considered for predicting DRAM-caused node unavailability (DCNU), emphasizing the importance of spatio-temporal CE features. Furthermore, in \cite{Li_intel_bit_2022}, specific error bit patterns are discussed and correlated with DRAM UEs. Rule-based error bit pattern indicators are developed for DRAM failure prediction across different manufacturers and part numbers, aligning with the ECC design of contemporary Intel Skylake and Cascade Lake servers. In addition, HiMFP framework \cite{huawei_2023_dsn} advocates a hierarchical system-level approach to memory failure prediction, using error bits features. \textit{However, the intrinsic distributions of error bits, specifically in Data pins (DQ) and beat, remain unexplored in above literature. Delving into these distributions is crucial for understanding the correlation between CE and UE.}

In this paper, we present an in-depth correlative analysis between CE and UE, specifically focusing on the spatio-temporal distribution of error bits. We also investigate latent patterns of error bits from CE to UE on the ECC of contemporary Intel servers. Our primary goal of analysis is to enhance memory failure prediction based on various DRAM errors and system configurations. Finally, machine learning models are implemented to leverage spatio-temporal error bits for memory failure prediction.

\textbf{The key contributions of the paper are as follow:}
\begin{itemize}
    \item We analyze error bits patterns generated from DIMM manufacturer and part number, and construct novel temporal risky CE indicators for UE prediction.
    \item We conduct the first in-depth correlative analysis between error bits and UE, specifically during DRAM read/write in Data pins (DQ) and beat. In addition, micro-level faults in the memory subsystem and system configurations are further correlated with UE occurrences. 
    \item We design ML-based failure prediction algorithms, based on the statistical insights from our analyses. Through evaluations using real-world data from a large-scale data center, our proposed error bits features have demonstrated the ability to capture latent patterns within the ECC of contemporary Intel servers, significantly improving UE prediction. When compared to the state-of-the-art algorithm \cite{Li_intel_bit_2022}, our approach achieves up to an 15\% improvement in F1-score for UE prediction, resulting in approximately a 59\% Virtual Machine Reduction Rate (VIRR) in our data centers.
\end{itemize}

The remainder of this paper is organized as follows: Section~\ref{sec:backgroundandmotivation} provides the background of our work. Section~\ref{sec:dataset} discussed the dataset employed in our data analysis. In Section~\ref{sec:Problem Formulation and performance-measures}, we formulate the the problem and define the performance measurements. Section~\ref{sec:risky CE patterns} introduces error bit pattern indicators for UE prediction. Section~\ref{sec:correlative analysis} presents an correlative study on UE. Section~\ref{sec:failure_prediction} demonstrates machine learning techniques for memory failure prediction. Experimental results are shown in Section~\ref{sec:result}. Section~\ref{sec:conclusion} concludes this paper.

\section{Background}
\label{sec:backgroundandmotivation}

\subsection{Terminology}
\label{ssec:terminology}
A \textit{fault} serves as the underlying cause of an error in DRAM, and it can be caused by various factors such as particle impacts, cosmic rays or defects. 

An \textit{error} refers to the situation in which a DIMM provides data to the memory controller that is inconsistent with the ECC \cite{SECDED,chipkill1997,intel_sddc,supermicro_sddc}, resulting from an active fault. Depending on ECC's capability to correct them, memory errors can be classified into correctable errors (CEs) and uncorrectable errors (UEs) \cite{vilas_systematic_study}. Two specific types of UEs are well-studied in prior literature\cite{giurgiu_predicting_2017}. 1) \textit{sudden UE}: UEs caused by some component faults that instantly corrupt data, and 2) \textit{predictable UE}: UEs that initially manifest as correctable errors but eventually escalate into UEs. A sudden UE typically has no CEs before it occurs, while a predictable UE can be predicted using CEs with failure prediction algorithms.

\begin{figure*}[t]
\centering

\vspace{-0.5cm}
\centering{\includegraphics[width=1\linewidth,keepaspectratio]{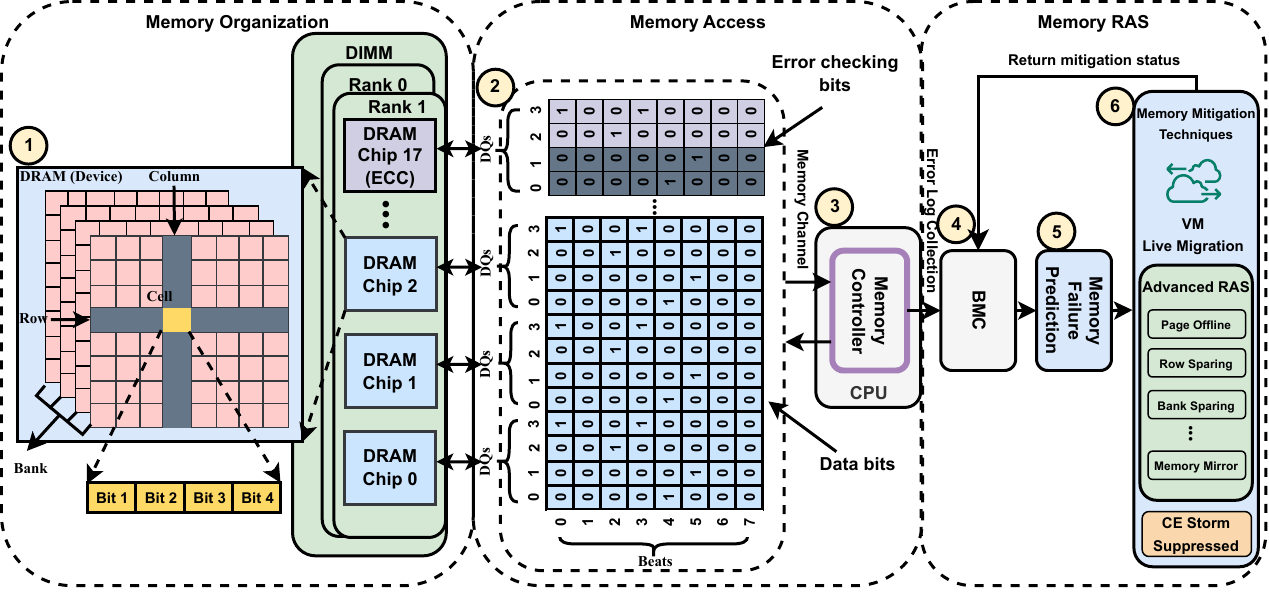}} 

\caption{Illustration of Memory Failure Prediction Framework.}  
\label{fig:A Framework of Memory Failure Prediction}
\vspace{-0.6cm}
\end{figure*}

\subsection{Memory Organization and Access}
\label{sec:memory Access}
Figure~\ref{fig:A Framework of Memory Failure Prediction} illustrates a framework of memory subsystem organization, memory access and memory RAS. The memory system is hierarchical in Figure~\ref{fig:A Framework of Memory Failure Prediction}(1): A DIMM rank is composed of several DRAM chips that form banks of two-dimensional arrays. Each bank is organized into rows and columns, and each addressable unit indexed by rows and columns is a memory cell containing a 4-bit word in the x4 DRAM device. Data flow in this architecture is transmitted from the cell to memory controller, which can generally detect and correct CEs via channels. Figure~\ref{fig:A Framework of Memory Failure Prediction}(2) depicts the transmission process of x4 DRAM Double Data Rate 4 (DDR4) chips via DQs. Upon initiating a data request, 8 beats each with 72 bits (64 data bits and 8 ECC bits) including ECC error codes are transferred to memory controller via DQ wires. Implementing the contemporary ECC \cite{Li_intel_bit_2022,supermicro_sddc}, 72-bit data are spread across 18 DRAM chips, allowing the memory controller to detect and correct them with ECC in Figure~\ref{fig:A Framework of Memory Failure Prediction}(3). Note that ECC checking bits addresses are decoded to locate specific errors in DQs and beats. Then, all these logs including error detection and correction, events, and memory specifications are archived in Baseboard Management Controller (BMC)\footnote{BMC is a delicated processor integrated into server's motherboard, tasked with monitoring the physical state of a computer, network server, or other hardware device.} in Figure~\ref{fig:A Framework of Memory Failure Prediction}(4). Among previous works \cite{du_memory_2018,du2020-intel,boixaderas_cost-aware_2020,Yu_drampakdd_2021,Du_nofreepredictor_2021,alibaba_2022_node_prediction,Li_intel_bit_2022,huawei_2023_dsn}, error bits in a cell have not been extensively examined. In our work, we conduct the first in-depth correlative analysis between error bits and UE in the field, to unveil the latent patterns of memory UEs.

\subsection{DRAM RAS Techniques }
\label{ssec:RAS}
DRAM subsystems are typically protected by RAS features in Figure~\ref{fig:A Framework of Memory Failure Prediction}(6). Proactive early VM live migrations can greatly reduce VM interruptions by moving VMs without service interruption. The CE storm suppressed mechanism helps avoid service degradation caused by CE storm\footnote{\label{ce_storm}CE interruptions repeatedly occur multiple times, e.g., 10 times.}. Advanced RAS techniques are designed to protect server-grade machines include the avoidance of fault regions. On the hardware technologies, sparing mechanisms are employed, such as bit sparing (e.g., Partial Cache Line Sparing (PCLS)\cite{Partial_cache_line_sparing}), row/column sparing (e.g., Post Package Repair (PPR) \cite{spare_row_column}), bank/chip sparing (e.g., Intel’s Adaptive Double Device Data Correction (ADDDC) \cite{du_page_offline_2021,spare_bank_chip}), etc. On the software-sparing mechanisms, such as the page offlining in operating systems, can also be applied to avoid memory errors \cite{du_page_offline_2021,Du_pageoffline_2019,mem_page_retirement}. However, these techniques often require higher redundancy and entail additional overhead, which can potentially impact system performance. Hence, these techniques cannot be universally adaptable across all machines. Utilizing memory failure prediction allows for the prediction of UEs and the activation of corresponding mitigation techniques based on specific use cases.

\section{Dataset}
\label{sec:dataset}
Our dataset was obtained from the Baseboard Management Controller (BMC) of a large-scale datacenter, which includes system configuration, Machine Check Exception (MCE) log \cite{Kleen2010mcelogM}, and memory events. We focus on DIMMs with CEs, excluding those with sudden UEs from our datasets due to a lack of prediction information. The MCE log records both CE and UE, providing details about memory error addresses (e.g., rank, bank, column) and DIMM specifications (e.g., manufacturer, capacity). We examined error logs from approximately 200,000 servers with Intel Skylake (Launched in 2017), Cascade Lake (Launched in 2019), Cooperlake (Launched in 2020) and Icelake (Launched in 2021) architectures in the datacenter.

Table~\ref{tab:data-description} provides an overview of the collected data. For the training set, we gathered over 80,000 Double Data Rate 4 (DDR4) DIMMs, spanning different manufacturers and part numbers, with CEs recorded from January to September 2022. Among them, we observed over 2,000 DIMMs with UEs, with 71\% of UE DIMMs having preceding CEs and 29\% are sudden UEs. Using a consistent collection approach, we prepared over 30,000 DIMMs for the test set from October to December 2022. This test set included over 1,000 DIMMs with UEs, with 67\% of UE DIMMs having preceding CEs and 33\% are sudden UEs. We conducted our correlative analysis and algorithm training based on the train set. The test set is reserved for final evaluation in Section~\ref{sec:result}.

\begin{table}[t]
    \centering
    \caption{Description of dataset.}
    \begin{tabular}{|c|c|c|c|c|}
    \hline
          \multirow{2}{*}{Dataset} & \multirow{2}{*}{Timespan}  & DIMMs  & DIMMs       \\
          & & with CEs& with UEs \\
          \hline
          Train set & 9 months  & $>$ 80,000 & $>$ 2,000    \\
         \hline
          Test set  & 3 months  & $>$ 30,000  & $>$ 1,000 \\ 
         \hline
    \end{tabular}
    
    \label{tab:data-description}
    \vspace{-0.5cm}
\end{table}

\section{Problem Formulation and Performance Measures}
\label{sec:Problem Formulation and performance-measures}

\begin{figure}[t]
\centering
\includegraphics[width=1\linewidth]{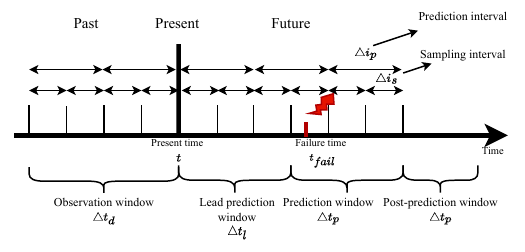}
\vspace{-0.7cm}
\caption{Failure prediction problem definition \cite{huawei_2023_dsn}.} 
\label{fig:failure-prediction-problem}
\vspace{-0.7cm}
\end{figure}

The failure prediction problem is formulated as a binary classification problem \cite{huawei_2023_dsn}. As illustrated in Figure~\ref{fig:failure-prediction-problem}, at present $t$, an algorithm observes historical data from an \textit{observation window} $\bigtriangleup t_{d}$ to predict failures within the prediction period $[t+\bigtriangleup t_{l}, t + \bigtriangleup t_{l} + \bigtriangleup t_{p}]$, where $\bigtriangleup t_{l}$ is a minimum time interval between the prediction and the failure. $\bigtriangleup t_{p}$ denotes the prediction interval. Online event samples are taken every $\bigtriangleup i_{s}$, e.g, CE events are logged every minute. Predictions run at 5-minute intervals  $\bigtriangleup i_{p}$. Observation and prediction windows are set at 5 days ($\bigtriangleup t_{d}$) and 30 days ($\bigtriangleup t_{p}$) respectively, enabling proactive measures. Note that these parameter settings were derived from an empirical analysis in the production environment. The \textit{lead prediction window} $\bigtriangleup t_{l} \in (0, 3h]$ varies based on production use cases. A True Positive (TP) is a correctly predicted failure within the prediction window, while a False Positive (FP) is an incorrect prediction. A failure without a prior alarm is a False Negative (FN), and a True Negative (TN) occurs when no failures are predicted or occur. We assess the algorithm using $Precision = \frac{TP}{TP + FP }$, $Recall = \frac{TP}{TP + FN }$ and $F1 = \frac{2\times Precision\times Recall}{Precision + Recall } $.

\begin{figure*}[htbp]
  \centering
  \begin{subfigure}{0.32\textwidth}
    \centering
    \includegraphics[width=\linewidth]{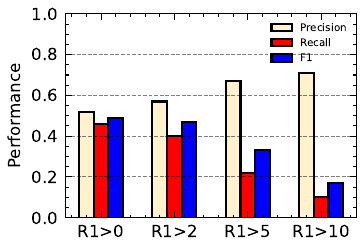}
    \vspace{0.01cm}
    \caption{R1: Risky CE number}
    \label{fig:risky_ce_number}
  \end{subfigure}%
  \begin{subfigure}{0.32\textwidth}
    \centering
    \includegraphics[width=\linewidth]{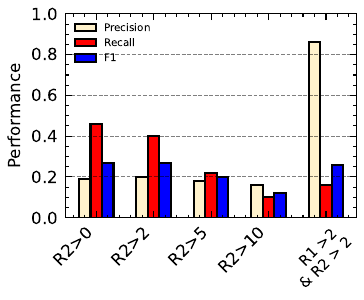}
    \vspace{-0.85cm}
    \caption{R2: Risky pattern number}
    \label{fig:risky_pattern_number}
  \end{subfigure}%
  \begin{subfigure}{0.32\textwidth}
    \centering
    \includegraphics[width=\linewidth]{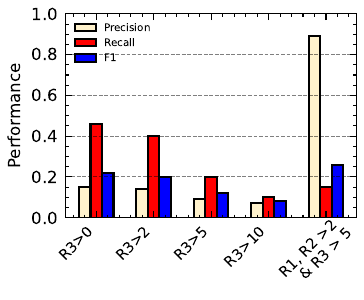}
    \vspace{-0.85cm}
    \caption{R3: Max risky patterns}
    \label{fig:max_risky_pattern}
  \end{subfigure}
  \caption{Performance analyses of risky CE patterns.}
  \label{Performance analyses of risky CE pattern.}
  \vspace{-0.5cm}
\end{figure*}

\textbf{\textit{VM Interruption Reduction Rate (VIRR)}}. Previous works \cite{boixaderas_cost-aware_2020,Du_nofreepredictor_2021,alibaba_2022_node_prediction,Li_intel_bit_2022,huawei_2023_dsn} have proposed cost-aware models to measure the benefits of memory failure prediction. In this work, we focus on \textit{VM Interruption Reduction Rate (VIRR)} \cite{huawei_2023_dsn} as it more accurately reflects the impact on customers. 

To understand VIRR, consider $V_{a}$ as the average number of VMs in a server. In a scenario devoid of prediction, the interruptions are defined as $V=V_{a}(TP + FN)$. Even though proactive VM live migrations can reduce VM interruptions without service interruption, a notable fraction of VMs may still experience cold migration, which generally interrupts VMs. This cold migration typically ensues when live migrations cannot be applied, either due to a paucity of resources or unforeseen failures. Given that cold migration is a prevalent strategy for both VM relocation and maintenance. The percentage of such migration is represented as $y_c$. Therefore, we define $V'_{1} = V_{a} \cdot y_{c}(TP + FP)$ as the number of VM interruptions arising from cold migrations initiated by positive failure predictions (TP + FP). On the other side, any missed failure predictions invariably escalate the interruptions, represented by $V'_{2} = V_{a}\cdot FN$. The overall interruptions after factoring in the prediction algorithm sum up to $V' = V'_{1} + V'_{2}$. The formula to measure VIRR is thus: $VIRR = \frac{V - V'}{V}$. Simplifying this give us $(1 - \frac{y_{c}}{precision}) \cdot recall$ as derived in \cite{huawei_2023_dsn}.

In real-world production environments, $y_{c}$ retains a positive value as VMs can be cold migrated due to the failure of live migration or memory recovery. If a model's precision dips below the percentage of cold migration ($precision < y_{c}$), the VIRR becomes negative, indicating an increase in VM interruptions. In contrast, models with high precision consistently yield a positive VIRR, and this is further amplified by the recall. Based on our observations in the production environment, we have defined $y_{c} = 0.1$ for our evaluation. Note that this value is already pessimistic, as the cloud infrastructure continues to expand, leading to a decrease in $y_{c}$ over time.

\section{Temporal Risky CE Pattern Indicators}
\label{sec:risky CE patterns}
According to a recent study by Intel \cite{Li_intel_bit_2022}, ECCs in modern Intel server platforms do not fully cover every potential errors from a single chip. Although Intel keeps the exact ECC algorithms confidential and undisclosed, they have provided some general information on error-bit patterns that can be fully correctable, partially correctable and potential risky in \cite{Li_intel_bit_2022,du_page_offline_2021,DDR5_bounded_fault}. For example, as shown in Figure~\ref{fig:A Framework of Memory Failure Prediction}(2), a DIMM with x4 DRAMs provides 32 error checking bits across 4 DQs and 8 beats during memory access. In a specific ECC outlined in \cite{DDR5_bounded_fault}, if all the actual erroneous bits are bounded within the half of the bitmap (highlighted in gray in the error checking bits), that error is guaranteed to be correctable. Otherwise, it is risky. More publicly available examples of error bit patterns can be found in \cite{Li_intel_bit_2022,du_page_offline_2021,DDR5_bounded_fault}. 

In this paper, we also obtain coarse-grained error-bit patterns, such as risky error bit patterns that are more likely to encounter UEs on contemporary Intel servers. CEs with risky error bit patterns are prone to evolve to UEs that cannot be corrected by the modern ECC algorithm \cite{Li_intel_bit_2022}. We introduce three temporal risky pattern indicators as follows:
\begin{itemize}
    \item \textbf{R1: Risky\_CE\_Cnt}: The number of unique CEs that match at least one risky error-bit pattern in a 24-hour period;
    \item \textbf{R2: Risky\_Pattern\_Cnt}: Total number of matched risky error-bit patterns in a 24-hour period; 
    \item \textbf{R3: Max\_Risky\_Pattern\_Cnt}: Maximum number of unique matched risky error-bit patterns counted in a 24-hour period;
\end{itemize}

While R1 is similar to the indicator in \cite{Li_intel_bit_2022}, R2 and R3 are novel pattern indicators proposed in this work. We compare the performance of these three indicators in Figure~\ref{Performance analyses of risky CE pattern.}. As shown in Figure~\ref{Performance analyses of risky CE pattern.}(a), when the count of risky CEs is greater than 0 (indicating at least one risky CE), it achieves 52\% precision, 46\% recall and 49\% F1-score on the training set. As the count of risky CEs increases, the precision also increases accordingly. However, the recall drops, indicating that most DIMMs with UE originate from a small number of risky CEs. This is intuitive that We evaluate the performance of R2 indicator in Figure~\ref{Performance analyses of risky CE pattern.}(b) and observe that its performance does not increase linearly as the count of matched risky patterns increases. However, when combining R1 $>$ 2 and R2 $>$ 2, the precision increases significantly to 86\%. On the other hand, individual R3 does not perform well on its own in Figure~\ref{Performance analyses of risky CE pattern.}(c). However, when combined with R1 and R2, it improves precision to the highest value of 89\%. Therefore, combining different pattern indicators can effectively enhance performance, which motivates us to use machine learning to integrate all indicators and correlated features, aiming to further improve UE prediction. Additionally, the risky patterns originate from the distribution of error bits in Data pins (DQs) and beats. We delve deeper into investigating the spatial and temporal distribution of error bits in DQs and beats in Section~\ref{ssec:Error Bits analysis}.

\textbf{Finding 1.} \textit{The performance of an individual risky CE pattern is limited. However, the proper combination of risky CE pattern indicators can significantly improve the results, particularly precision. }

\section{Correlative Analysis between Uncorrectable Error and Various Factors}
\label{sec:correlative analysis}
We start with the high-level of correlative study between UE and various factors. Specifically, we investigate the relationship among error bits, DRAM faults, and system configurations to gain insights into their influence on UE occurrences. This analysis is essential for identifying relevant features that can be used for model training and failure prediction as outlined in Section~\ref{sec:failure_prediction}. Our methodology follows a similar approach to previous studies \cite{Meza_RevisitingError_2015,Sridharan_memory_error_modern_system,srds_in-Depth_MEM}. We employ a calculation method named as relative UE rate, where DIMMs are grouped based on specific characteristics (e.g., server age), and the fraction of DIMMs experiencing UEs is determined. The relative UE rates are normalized within the range [0, 1], enabling us to observe trends, compare rates, and finally extract important features for UE prediction.

\subsection{Correlative Analysis Between Error Bits and UE}
\label{ssec:Error Bits analysis}

\begin{figure}[!h]
\centering
\vspace{-0.3cm}
\includegraphics[keepaspectratio]{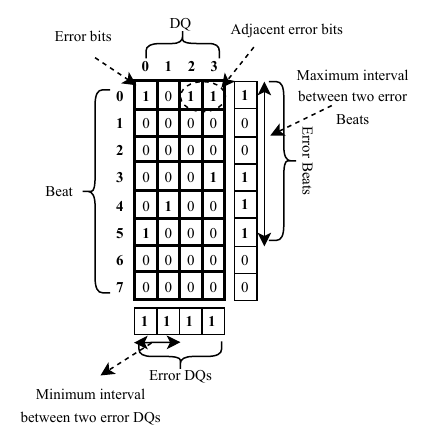}
\vspace{-0.3cm}
\caption{Spatial correlations of error bits in DQs and Beats.} 
\label{fig:bit_feature}
\vspace{-0.5cm}
\end{figure}

\begin{figure}[!h]
\centering
\includegraphics[keepaspectratio]{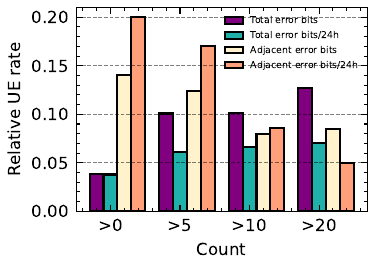}
\vspace{-0.3cm}
\caption{Error bits analysis.} 
\label{fig:Error bits analysis}
\vspace{-0.5cm}
\end{figure}

We first examine the relative UE rate based on characteristic of error bits. To quantify this, we first calculate the total number of error bits and the number of adjacent error bits within a single CE event. For a specific memory access, Figure~\ref{fig:bit_feature} visualizes bitmap of error bits occurring in four DQs and four beats. In this example, there are total six error bits and one pair of adjacent error bits. Figure~\ref{fig:Error bits analysis} illustrates the correlation between the total number of error bits and the UE rate. As the count of error bits increases, the UE rate generally rises. However, the overall relative UE rates remain relatively low.

A finding emerges when comparing the relevance of adjacent error bits with the total number of error bits. The occurrence of adjacent error bits within a specific range, such as greater than 0 or 5, is more strongly associated with the occurrence of UEs. This implies that even a small number of adjacent bits can have a risk for UE occurrence.

\textbf{Finding 2.} \textit{In terms of UE occurrence, the total number of error bits exhibits weaker correlation compared to adjacent error bits. Even a small number of adjacent bits can lead to UE occurrence.}

We then investigate the spatial distribution of error bits in DQs and beats. As shown in Figure~\ref{fig:bit_feature}, we have calculated the number of error DQs and beats, which yields four error DQs and four error beats. We also examine other key features such as the interval between error DQs and the interval between error beats.

\begin{figure*}[tbh]
\centering
\includegraphics[width=1\linewidth, keepaspectratio]{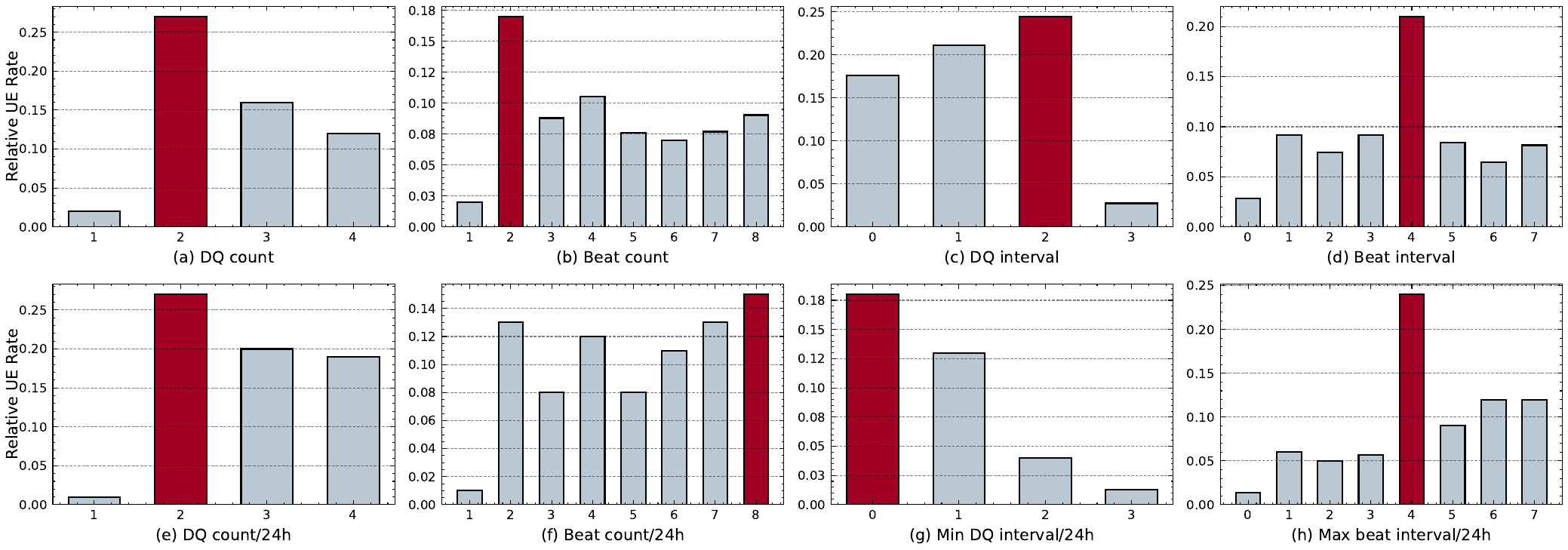}
\caption{Analyses of spatial and temporal error bits: Highlighting the highest rate with red bar.} 
\label{fig:spatio_temporal_Error_bits_analysis}
\vspace{-0.4cm}
\end{figure*}

The correlative analysis of these spatial features is presented in Figure~\ref{fig:spatio_temporal_Error_bits_analysis}. In Figure~\ref{fig:spatio_temporal_Error_bits_analysis}(a), error DQs with two, three, or four generally exhibit higher UE rates compared to those CEs with only one error DQ. Similarly, Figure~\ref{fig:spatio_temporal_Error_bits_analysis}(b) indicates that multiple error beats have higher UE rates compared to one error beat. Error bits occurring in more than one DQ and beat are more likely to encounter UEs. Additionally, our analysis reveals an important observation regarding the interval between error DQs and beats. Specifically, we found that the error DQs interval of three exhibit a relatively lower UE rate compared to other intervals. On the other hand, beats interval of four have the highest UE rate compared to other intervals. These insights highlight the importance of considering the specific intervals between error DQs and beats in understanding the occurrence of UE.

\begin{figure}[tbh]
\centering
\vspace{-0.4cm}
\includegraphics[keepaspectratio]{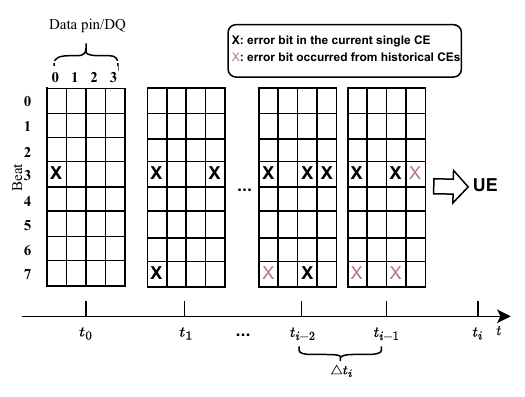}
\vspace{-1cm}
\caption{An example of error bits propagation.}
\label{fig:bit propagation}
\vspace{-0.6cm}
\end{figure}

In addition to spatial correlative analysis of error bits in DQs and beats, we incorporate temporal information into these features. As shown in Figure~\ref{fig:bit propagation}, error bits are propagated through DQs and beats over time in $t$, which can be increased from a single one in $t_{0}$ to multiple bits spanning across DQs and beats, eventually lead to UE. Note that error bits of UEs are typically unknown, since they are not correctable, typically leading to service down without logging error addresses. In terms of spatial distribution, a single CE event in $t_{i-1}$ may involve 2 error DQs within the same beat. However, in the case of multiple CEs within an interval $\bigtriangleup t_{i}$, there could be three error DQs spanning across 2 beats. To capture these spatio-temporal bits patterns, we calculate statistical features such as \textit{Sum}, \textit{Maximum}, \textit{Minimum}, \textit{Average} and \textit{Standard deviation} of error bits in DQs and beats based on all CE events within the aggregation window $\bigtriangleup t_{i}$. Additionally, we analyze spatio-temporal features of DQ and beat counts and intervals within a 24-hour aggregation window in Figure~\ref{fig:spatio_temporal_Error_bits_analysis}(e)-(h). While the relative UE rates for temporal error DQs and beats in Figure~\ref{fig:spatio_temporal_Error_bits_analysis}(e) and (f) are vary with Figure~\ref{fig:spatio_temporal_Error_bits_analysis}(a) and (b) respectively, the consistent trend remains that one error DQ or beat has a lower relative UE rate compared to multiple error DQs and beats. Furthermore, the minimum error DQ interval and the maximum error beat interval within 24 hours exhibit different relative UE rates in Figure~\ref{fig:spatio_temporal_Error_bits_analysis} (g) and (h). Among all error DQ intervals, the interval of 3 consistently exhibits the lowest UE rate. On the other hand, among all beat intervals, the interval of 4 demonstrates the highest UE rate.

\textbf{Finding 3.} \textit{Our analyses reveal that both spatial and temporal error bits in DQs and beats play a significant role in distinguishing UE occurrences. This finding suggests that these features can serve as important indicators for UE prediction.}

Therefore, we generate both spatial error bits features in a single CE and spatio-temporal error bits features across multiple CE events for UE prediction. Even features with relatively low UE rates may still contribute significantly when utilized in conjunction with machine learning techniques for UE prediction. We conduct feature selection and UE prediction based on machine learning in Section~\ref{sec:failure_prediction}.

\subsection{Correlative Analysis Between DRAM Faults and UE}
\label{sec:DRAM faults analysis}
CEs can originate from various components within the memory subsystem, as depicted in Figure~\ref{fig:A Framework of Memory Failure Prediction}(1). To examine the impact of different component faults on memory failure, which ultimately leads to the generation of error bits during memory access as shown in Figure~\ref{fig:A Framework of Memory Failure Prediction}(2). We consider DIMM-level of components' faults from cell, column, row, bank, device and rank respectively. If the number of CEs repeated in the same cell reaches a predefined threshold $\theta_{cell}$, it refers to \textbf{Cell fault}. If CEs scattered along in a row and a column reaches $\theta_{row}$ and $\theta_{column}$, they are \textbf{Row fault} and \textbf{Column fault} respectively. \textbf{Bank fault} refers to the case where row faults and column faults both are greater than $\theta_{bank}$ in the same bank. More than $\theta_{device}$ of unique bank faults occurred in a device indicates \textbf{Device fault}. \textbf{Rank fault} represents that deveice faults reach a predefined threshold $\theta_{rank}$ in the same rank. We defined $\{\theta_{cell},\theta_{row}, \theta_{column},\theta_{Device}, \theta_{Rank}\} = 2$ and  $\theta_{bank} = 3$ in our analyses.

\begin{figure}[tbh]
\centering
\vspace{-0.3cm}
\includegraphics[width=1\linewidth, keepaspectratio]{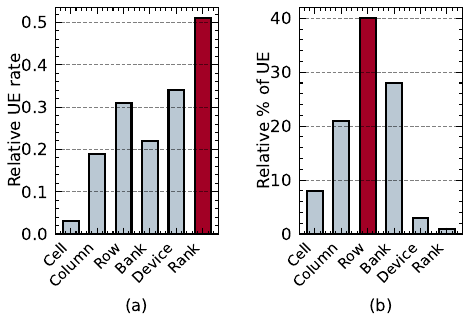}
\caption{Micro-level components' fault analysis.} 
\label{fig:micro_level_fault_analysis}
\vspace{-0.3cm}
\end{figure}

\begin{figure*}[tbh]
\centering
\includegraphics[width=1\linewidth, keepaspectratio]{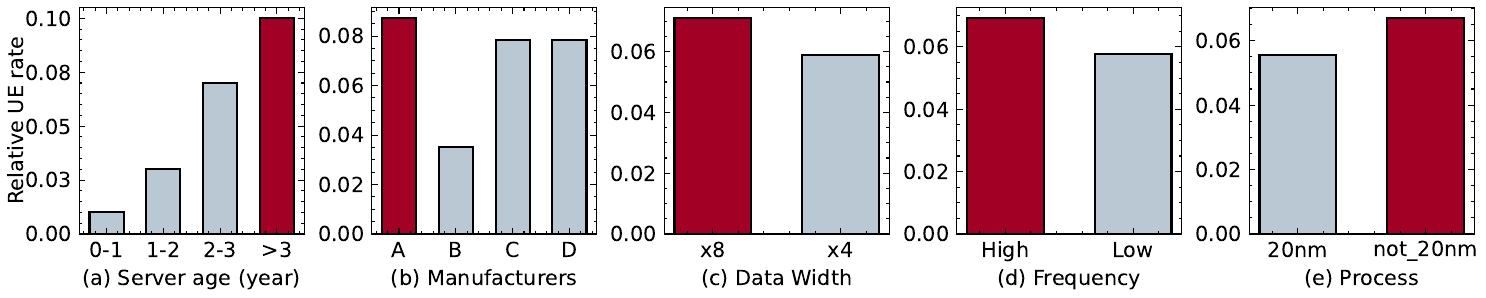}
\vspace{-0.5cm}
\caption{System configurations analyses.} 
\label{fig:dram_configuration_analysis}
\vspace{-0.4cm}
\end{figure*}

We first examine each component fault by excluding the higher-level faults. For example, As shown in Figure~\ref{fig:micro_level_fault_analysis}, Cell faults exhibit a UE rate of less than 0.2. However, when cell faults accumulate and propagate to higher levels of DRAM components, relatively 0.31 of UE rate associated with row faults (excluding column and higher-level faults, such as bank, device and rank), and 0.22 of UE rate associated with bank faults (excluding device and rank faults). We also visit relative percentage of UE in each component fault. Although Device and Rank faults have a higher relative UE rate, the proportion of UEs associated with these faults is relatively small compared to Row and Bank faults.

\textbf{Finding 4.} \textit{While higher-level faults may have a higher likelihood of causing UEs, Row and Bank faults account for the majority of UEs in the system. This emphasizes the importance of addressing and mitigating Row and Bank faults to improve the overall reliability and performance of the memory subsystem.}

\subsection{Correlative Analysis Between System Configuration and UE }
\label{sec:DRAM System Configuration analysis}

In our study, we examine the correlation between system configurations and UEs. We first analyze the server age, and our findings align with our conjecture that older servers are more likely to experience memory failures. Figure~\ref{fig:dram_configuration_analysis}(a) shows that servers with more than 2 years of age have a higher UE rate. 

Furthermore, we investigated various DRAM hardware configurations, including manufacturer, capacity, device data width, frequency, and process in Figure~\ref{fig:dram_configuration_analysis}. To protect the confidentiality of manufacturer names, we anonymized them as manufacturers A to D, representing the four major DIMM manufacturers in our data centers. Different manufacturers exhibited varying UE rates, potentially due to differences in DIMM processes.

We also observe that DIMMs with x8 bit width have a higher relative UE rate compared to those with x4 bit width. This difference may be attributed to variations in memory access and ECC correction. Additionally, higher DRAM frequency generally correlates with higher relative UE rates. We further examine the DRAM process, categorizing them as either 20nm or not (as the exact processes of 1ynm, 1xnm, and 1znm are proprietary information). The not 20nm process category shows a higher relative UE rate.

The capacity of the DIMM did not significantly impact the UE rate in our study. 

\textbf{Finding 5.} \textit{The UE rate varies across server age, manufacturers, data width, frequency and process, while we did not observe significant differences in the UE rate based on the capacity of the DIMM.}

These attributes, including server age, manufacturer, data width, frequency, and process, can be valuable for failure prediction in Section~\ref{sec:failure_prediction}.

\section{Failure Prediction}
\label{sec:failure_prediction}
In this section, we design memory failure prediction based on pattern indicators (Section~\ref{sec:risky CE patterns}) and correlative analysis between UE and various factors (Section~\ref{sec:correlative analysis}).

We develop failure prediction mainly using machine learning techniques, e.g., Random Forest\cite{alibaba_2022_node_prediction}, XGboost\cite{alibaba_2022_node_prediction}, LightGBM\cite{srds_in-Depth_MEM} and AdaUboost \cite{Du_nofreepredictor_2021}, since these ensemble learning techniques have been widely used in previous memory failure prediction literature \cite{alibaba_workload-aware_2021,Yu_drampakdd_2021,Du_nofreepredictor_2021,boixaderas_cost-aware_2020,giurgiu_predicting_2017} due to their fast learning and good performance. The experimental results of these models are presented in Section~\ref{sec:result}.

\textbf{Labeling method}: Our prediction framework categorizes samples into two classes: Positive and Negative. DIMMs expected to encounter at least one UE within the prediction window are categorized as \textbf{Positive}, whereas those not expected to experience any UE are termed \textbf{Negative}.

Positive samples are labeled based on the time interval $t_{i}$ between a CE and its subsequent UE. Selected intervals for $t_i$ include 6 hours, 24 hours, 72 hours, 120 hours, 1 month, and a DIMM's entire lifetime. CEs that fall within the 0 to $t_{i}$ interval preceding a UE are marked as \textbf{Positive smaples}. CE events outside this period are excluded to prevent mislabeling. All CE events from healthy DIMMs are labeled as \textbf{Negative samples}. However, our training data experiences from class imbalance, we employ over-sampling strategies for positive samples, ensuring models adequately address both classes.

\textbf{Feature generation.} We categorize features into six groups including: 
\begin{itemize}
    \item \textit{Static Features} describe DIMM characteristics studied in Section~\ref{sec:DRAM System Configuration analysis} including server age, manufacturers, data width, frequency and chip process.
    \item \textit{CE error rate} refer to the number of CEs and their occurrence frequency, e.g., error counts of all CEs within the predefined time.
    \item \textit{DQ-Beat Error Bits} features refer to the spatial and temporal distribution of error bits in DQs and beats, as discussed in Section~\ref{ssec:Error Bits analysis}.
    \item \textit{Error bit Patterns} features are derived from three risky CE pattern indicators described in Section~\ref{sec:risky CE patterns}.
    \item \textit{Fault Counts} refers to the cumulative number of components' faults (cell, row, column, bank, device and rank) within $t_{i}$, derived from study in Section~\ref{sec:DRAM faults analysis}.
    \item \textit{Memory Events} refers to CE storm\textsuperscript{\ref{ce_storm}}, CE overflow\footnote{CE counts reach am initial overflow threshold.}, CE storm suppressed notification\footnote{The mechanism suppresses and notifies if a CE storm occurs several times in the same DIMM.}, etc, which indicate the unhealthy status of memory.
\end{itemize}

Totally, six groups of features are constructed as input of machine learning approaches. We select the best features using \textit{Pearson correlation}, \textit{Random Forest} and \textit{LightGBM} in Section~\ref{sec:result}. 

\section{Result}
\label{sec:result}
After empirical experiments on the training set, we explored the parameter $t_{i}$ ranging from 1 minute to 5 days. The output probability threshold was set to 0.3, since it can achieve the best VIRR with a predefined $y_{c}=0.1$. To evaluate the importance of designed features, three feature selection approaches are implemented in Table~\ref{tab:feature improtance ranking}. Among the top five important features identified by these approaches, four out of five are related to error bits, highlighting the significance of error bits in predicting UEs. Notably, \textit{Minmum error DQ interval} consistently ranked as the most important feature across all approaches. To determine the best feature set for algorithm training, we employed recursive feature elimination and feature importance ranking. Table~\ref{tab:ml-performance} displays the results, demonstrating that LightGBM outperformed other machine learning techniques with a F1-score of 0.64 on the test set. Consequently, we selected LightGBM for further analysis in our study.

\begin{table}[htb]
    \centering
    \caption{Rankings of the top five important features.}
    \resizebox{\columnwidth}{!}{
    \begin{tabular}{ |l| c|c| c|}
    \hline
           Rank & Pearson  & Random Forest & LightGBM\\
          \hline
          1  & \textbf{\textcolor{violet}{Min\_DQ\_interval}}  & \textbf{\textcolor{violet}{Min\_DQ\_interval}} & \textbf{\textcolor{violet}{Min\_DQ\_interval}}\\
         \hline
          2  & \textbf{\textcolor{violet}{Max\_beat\_interval}} & \textbf{\textcolor{violet}{Error\_DQ\_counts\_24h}} & Fault(Cell)\\
         \hline
          3  &  \textbf{\textcolor{violet}{Risky\_CE\_Cnt}} & CE overflow & \textbf{\textcolor{violet}{Risky\_CE\_Cnt}}\\ 
         \hline
         4   &  \textbf{\textcolor{violet}{Risky\_Pattern\_Cnt}} & \textbf{\textcolor{violet}{Max\_adjacent\_bits\_24h}} & \textbf{\textcolor{violet}{Risky\_Pattern\_Cnt}}\\ 
         \hline
         5   &  Fault(Row) & \textbf{\textcolor{violet}{Error\_beat\_Cnt}} & \textbf{\textcolor{violet}{Error\_DQ\_counts\_24h}}\\ 
         \hline
    \end{tabular}
    }
    \label{tab:feature improtance ranking}
\vspace{-0.3cm}
\end{table}

\begin{table}[t]
    \centering
    \caption{Performance of ML algorithms.}
    \begin{tabular}{ |l|l|l|l|}
    \hline
           Algorithms & Precision  & Recall & F1-Score\\
          \hline
         Random Forest      & 0.63  & 0.62 & 0.63\\
          \hline
          XGBoost  & 0.54  & 0.67 & 0.59\\
          \hline
           AdaUboost     & 0.54  & 0.78 & 0.64\\
          \hline
          LightGBM     & \textbf{0.53}  & \textbf{0.82} & \textbf{0.64}\\
          \hline
    \end{tabular}
    
    \label{tab:ml-performance}
    \vspace{-0.3cm}
\end{table}

\textbf{Comparison with existing approaches.} We further evaluate the significant of our proposed error bits features by comparing with existing the state-of-the-art approaches presented in \cite{Li_intel_bit_2022}. Specifically, we reproduced their rule-based approaches as discussed in Section~\ref{sec:risky CE patterns} and apply the same experimental setup on our dataset. Note
that the approaches in \cite{Li_intel_bit_2022} are designed with various part numbers of manufacturers, but the detail was not disclosed in their work. We evaluated their approaches without differentiating the part numbers. The results in Table~\ref{tab:Comparison with Existing approaches} demonstrate that our approach significantly achieves higher F1-score of 0.64 by including all features. In addition, our algorithm still achieves relatively better performance by excluding the error bits patterns features, which indicates the superior of error bits features in UE prediction. By excluding both error bits and pattern features, algorithm cannot perform well, which further prove the significant of error bits information for UE prediction.

\textbf{Finding 6.} \textit{The inclusion of error bits features significantly enhances UE prediction performance, even without knowledge of the error bits patterns. This alludes that the latent patterns of error bits can be predicted using spatio-temporal error bits features.}
\begin{table}[t]
    \centering
    \caption{Comparison with existing approaches.}
    \resizebox{\columnwidth}{!}{
    \begin{tabular}{ |l|l|l|l|}
    \hline
            Algorithms & Precision  & Recall & F1-Score\\
          \hline
            Risky\_CE Pattern  & 0.53  & 0.46 & 0.49\\
         
         Risky\_CE Pattern $\wedge$ Column       & 0.68  & 0.10 & 0.17\\
          
          Risky\_CE Pattern $\wedge$ Bank   & 0.84  & 0.11 & 0.19\\
          \hline
          Ours (Excluding error bits and patterns)    &  0.30  & 0.51 & 0.38 \\
          
          Ours (Excluding patterns)  & 0.45  & 0.74 & 0.56\\
          
          Ours (All features)      & \textbf{0.53}  & \textbf{0.82} & \textbf{0.64}\\
          \hline
    \end{tabular}
    }
    \label{tab:Comparison with Existing approaches}
    \vspace{-0.5cm}
\end{table}

\textbf{Lead time.} In Table~\ref{tab:lead-time}, we also examine the prediction results for three lead times. The lead time refers to the duration between the prediction time and the expected occurrence of a failure. Depending on the memory mitigation techniques, these lead times can vary. For instance, in a 15-minute lead time allows VM migration to a backup system and the deployment of advanced RAS techniques to prevent UE incidents. With a 1-hour lead time, VM migration may span up to an hour due to the workload involved, and failing machines can be localized and replaced with the corresponding DIMM. The VM Interruption Reduction Rate (VIRR) discussed in Section~\ref{sec:Problem Formulation and performance-measures} is estimated for these lead times. In our datacenters, we take into consideration a 15-minute lead time, which results in a reduction of approximately 59\% in VM interruptions caused by UEs.

\begin{table}[thb]
    \vspace{-0.3cm}
    \centering
    \caption{Performance in different lead times.}
    \begin{tabular}{ |c |c|c|c|c| }
        \hline
               Lead time & Precision & Recall & F1-Score & VIRR \\
               \hline
              1s & 0.53 & 0.82 & 0.64 &  0.67\\
               \hline
              15m & 0.46 & 0.75 & 0.57 & 0.59\\
             \hline
              1h & 0.36 & 0.45 & 0.40 & 0.33\\ 
             \hline
        \end{tabular}
        
        \label{tab:lead-time}
        \vspace{-0.5cm}
\end{table}

\section{Conclusion}
\label{sec:conclusion}
We present an in-depth correlative analysis on uncorrectable errors with various factors, particularly focusing on spatio-temporal error bits information of CEs. We report 6 findings from our analyses and failure prediction studies. Through evaluations using real-world datasets, we demonstrate that our approach significantly improves prediction performance by 15\% in F1-score compared to the state-of-the-art algorithms. Overall, it can reduce VM interruptions by around 59\% VIRR in the datacenter. In the future, we plan to extend our algorithm to include servers from different manufacturers' platforms, particularly focusing on the comparisons of Chipkill and non-Chipkill ECC servers. 
 
\section*{Acknowledgement}
We thank the anonymous reviewers from ICCAD'23 for their great comments.

\bibliographystyle{IEEEtran}
\bibliography{bibliography}

\end{document}